\documentclass[aps,prb,twocolumn,floatfix]{revtex4}
\usepackage{amsmath,amssymb,graphics,epsfig,epstopdf,color,verbatim,ulem,braket,tabularx,float}
\usepackage{makecell}
\usepackage{graphicx}
\usepackage{dcolumn}
\usepackage{bm}
\usepackage{times}
\usepackage{color}
\usepackage{appendix}

\begin{document}
\bibliographystyle{apsrev}
\title{Large anomalous Hall effect in layered antiferromagnet Co$_{0.29}$TaS$_2$}

\author{Huan Wang}
\author{Jun-Fa Lin}
\author{Xiang-Yu Zeng}
\author{Xiao-Ping Ma}
\author{Jing Gong}
\author{Zheng-Yi Dai}
\author{Xiao-Yan Wang}
\author{Kun Han}
\author{Yi-Ting Wang}
\author{Tian-Long Xia}\email{tlxia@ruc.edu.cn}

\affiliation{Department of Physics, Renmin University of China, Beijing 100872, P. R. China}
\affiliation{Beijing Key Laboratory of Opto-electronic Functional Materials $\&$ Micro-nano Devices, Renmin University of China, Beijing 100872, P. R. China}

\date{\today}

\begin{abstract}
We present a study on the magnetization, anomalous Hall effect (AHE) and novel longitudinal resistivity in layered antiferromagnet Co$_{0.29}$TaS$_{2}$. Of particular interests in Co$_{0.29}$TaS$_{2}$ are abundant magnetic transitions, which show that the magnetic structures are tuned by temperature or magnetic field. With decreasing temperature, Co$_{0.29}$TaS$_{2}$ undergoes two transitions at T$_{t1}\sim$ 38.3 K and T$_{t2}\sim$ 24.3 K. Once the magnetic field is applied, another transition T$_{t3}\sim$ 34.3 K appears between 0.3 T and 5 T. At 2 K, an obvious ferromagnetic hysteresis loop within H$_{t1}\sim\pm$ 6.9 T is observed, which decreases with increasing temperature and eventually disappears at T$_{t2}$. Besides, Co$_{0.29}$TaS$_{2}$ displays step-like behavior as another magnetic transition around H$_{t2}\sim\pm$ 4 T, which exists until $\sim$ T$_{t1}$. These characteristic temperatures and magnetic fields mark complex magnetic phase transitions in Co$_{0.29}$TaS$_{2}$, which are also evidenced in transport results. Large AHE dominates in the Hall resistivity with the conspicuous value of R$_{s}$/R$_{0}\sim 10^{5}$, considering that the tiny net magnetization (0.0094$\mu_{B}$/Co) alone would not lead to this value, thus the contribution of Berry curvature is necessary. The longitudinal resistivity illustrates a prominent irreversible behavior within H$_{t1}$. The abrupt change at H$_{t2}$ below T$_{t1}$, corresponding to the step-like magnetic transitions, is also observed. Synergy between the magnetism and topological properties, both playing a crucial role, may be the key factor of large AHE in antiferromagnet, which also offers a new perspective in magnetic topological materials with the platform of Co$_{0.29}$TaS$_{2}$.

\end{abstract}
\maketitle

\section{Introduction}
In the past decades,  great efforts have been made on the study of anomalous Hall effect (AHE), one of the fascinating manifestations in magnetic materials. Two qualitatively different mechanisms are widely accepted as its origin, intrinsic contribution from the Berry curvature and extrinsic impurity scattering including skew-scattering and side-jump model\cite{AHE}. The AHE is traditionally considered to be proportional to the spontaneous magnetization, and therefore
is believed to arise exclusively in ferromagnet \cite{AHE} and vanish in antiferromagnet due to the cancellation of magnetic moments. However, based on symmetry considerations\cite{Mn3Ir-1, Mn3Ir-2, AFM-1, AFM-2, AFM-3, AFM-4}, it was predicted that AHE may be realized in certain noncollinear antiferromagnet or frustrated magnet, as a result of nonvanishing contribution of Berry curvature. These predictions are experimentally confirmed in materials Mn$_{3}$X (X=Ge, Sn, Ir)\cite{Mn3Ge,Mn3Sn,Mn3X,Mn3Ir-1,Mn3Ir-2},Mn$_{5}$Si$_{3}$\cite{Mn5Si3}, GdPtBi\cite{GdPtBi}, Nd$_2$Mo$_2$O$_7$\cite{NdMoO}, Pr$_2$Ir$_2$O$_7$\cite{PrIrO}, \textit{etc.}, which deepens our understanding of the relation between topological nature and transport properties.

Recently, magnetic-atom-intercalated transition metal dichalcogenides M$_{x}$TaS$_{2}$ (M=V, Cr, Mn, Fe, Co), a unique system exhibiting rich electrical and magnetic behaviors\cite{series-1,series-2}, attract great interest and intense attention due to their 2D layered structure and large spin-orbit coupling. Depending on the intercalant element between TaS$_2$ layer, the materials M$_{x}$TaS$_{2}$ can be ferromagnetic (FM) such as V$_{1/3}$TaS$_{2}$\cite{VTa3S6}, Cr$_{1/3}$TaS$_{2}$\cite{CrTa3S6-1, CrTa3S6-2} and Mn$_{x}$TaS$_{2}$ (x = 1/4, 1/3)\cite{MnTa3S6, MnTa4S8}, or antiferromagnetic (AFM) such as Co$_{x}$TaS$_{2}$ (x = 0.22, 1/3)\cite{Co0.22TaS,Co0.33TaS}. Fe$_{x}$TaS$_{2}$ presents more interesting magnetic structures, FM for 0.2 $\leqslant$ x $\leqslant$ 0.4\cite{Fe0.25TaS-1,Fe0.25TaS-2,Fe0.28TaS-1,Fe0.28TaS-2} and AFM for x $>$ 0.4\cite{FexTaS-AFM}. Among these materials, novel magnetic properties including large magnetic anisotropy, giant anisotropic magnetoresistance and AHE have been reported. In this work, Co$_{0.29}$TaS$_2$, which exhibits complex magnetic structures and interesting magnetotransport properties, is studied in detail.

Co$_{0.29}$TaS$_{2}$ consists of magnetic Co atom intercalated between 2H-TaS$_{2}$ layer, which crystallizes in the hexagonal space group P6$_3$22. Its fascinating magnetic structures depend on temperature and field. With decreasing the temperature, Co$_{0.29}$TaS$_{2}$ undergoes two magnetic transitions at T$_{t1}\sim$ 38.3 K and T$_{t2}\sim$ 24.3 K. Meanwhile, with the intervention of magnetic field over the range of 0.3 T - 5 T, another transition at T$_{t3}\sim$ 34.3 K emerges. At 2 K, prominent hysteresis loop within H$_{t1}\sim\pm$ 6.9 T is observed, which decreases with increasing temperature and eventually disappears at T$_{t2}$. Besides, the magnetization displays a step-like behavior as another magnetic transition around H$_{t2}\sim\pm$ 4 T, which preserves until T$_{t1}$. The abundant magnetic transitions reveal tunable magnetic orders in Co$_{0.29}$TaS$_{2}$, resulting in novel transport properties. The striking finding is the remarkably large AHE with the value R$_{s}$/R$_{0}\sim 10^{5}$ despite a tiny net magnetization (0.0094 $\mu_{B}$ per Co). Consistent with magnetization results, the longitudinal resistivity shows a prominent irreversible behavior within H$_{t1}$. At the same time, the abrupt change at H$_{t2}$ below T$_{t1}$, corresponding to the step-like magnetic transitions, is also observed. These complex and interesting magnetic structures make Co$_{0.29}$TaS$_{2}$ an ideal 2D magnet to understand the interplay between magnetic structures, topology and the intrinsic mechanism of AHE in noncollinear antiferromagnet.

\section{Experiment}

\begin{figure}[htbp]
	\centering
 	\includegraphics[width=0.48\textwidth]{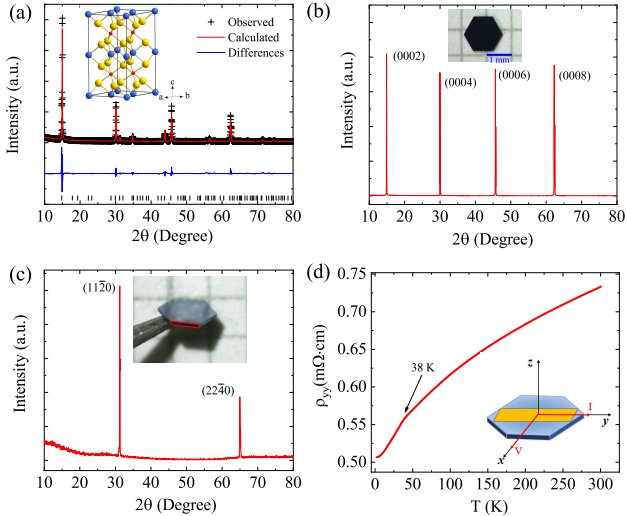}
	\caption{(a) Powder XRD patterns with refinement. Inset shows the crystal structure of Co$_{0.29}$TaS$_{2}$ with the space group P6$_3$22 (No. 182). R$_{wp}=13.07\%$. (b)(c) Single crystal XRD patterns of Co$_{0.29}$TaS$_{2}$ with different crystal surfaces. (d) The temperature-dependent resistivity from 2 K to 300 K. Inset shows the current applied along the edge.}
\end{figure}

\begin{figure}[htbp]
	\centering
	\includegraphics[width=0.48\textwidth]{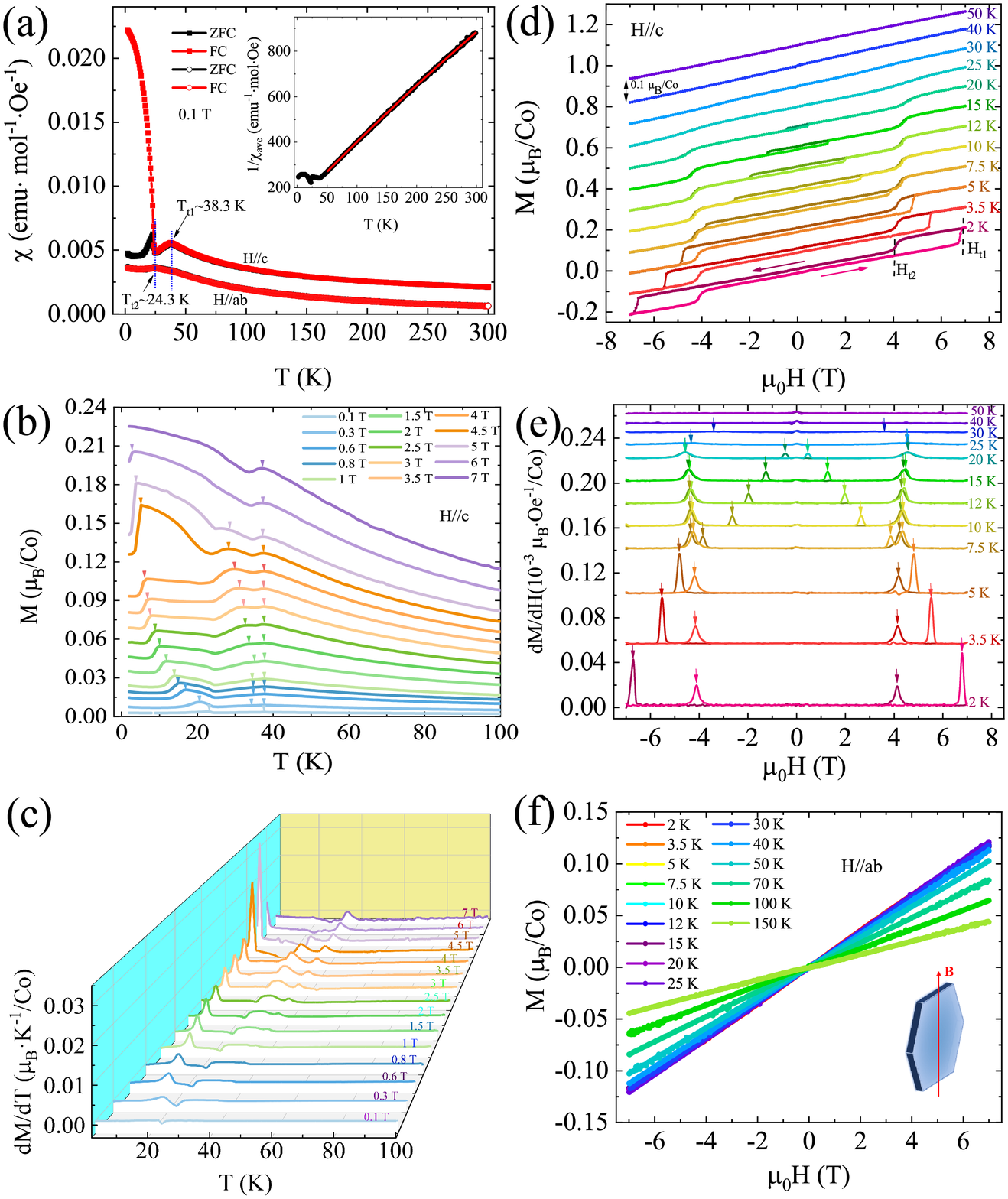}
	\caption{(a) Temperature dependence of the magnetic susceptibility along c axis and within ab plane, respectively. Inset shows the Curie-Weiss fit of inverse average susceptibility. (b) The temperature-dependent susceptibility under various fields along c axis. (c) The first-order derivative of the curves from (b). (d) The magnetization of Co$_{0.29}$TaS$_{2}$ with the magnetic field applied along c axis under different temperatures. For clarity, each curve except for the one at 2 K is shifted vertically with step of 0.1 $\mu_{B}$/Co. (e) The first-order derivative of curves in (d). (f) The magnetization with field applied within ab plane. Inset shows the sketch map that the magnetic field is perpendicular to the hexagonal edge and lies in the ab plane.}
\end{figure}

\begin{figure*}[htbp]
	\centering
	\includegraphics[width=\textwidth]{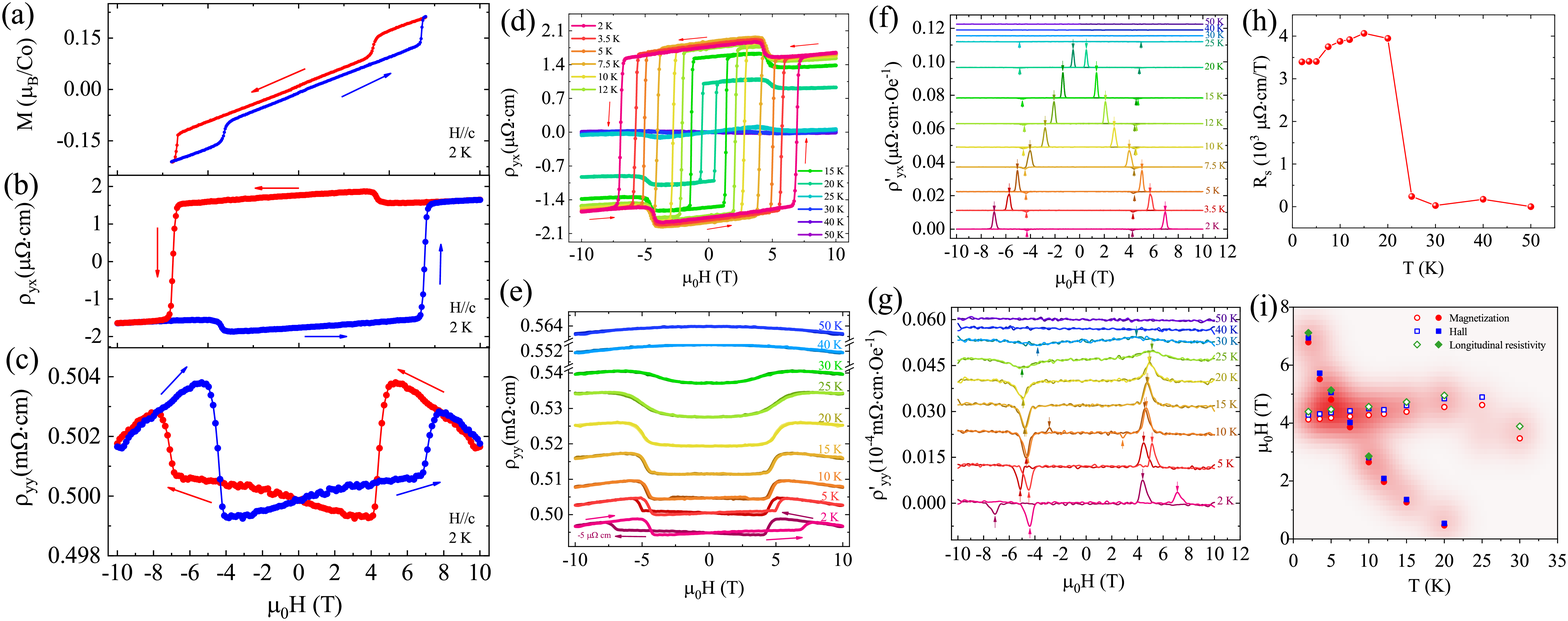}
	\caption{The field dependence of (a) magnetization with H up to $\pm 7$ T, (b) Hall resistivity and (c) longitudinal resistivity with H up to $\pm 10$ T along c axis at 2 K. (d) The Hall resistivity versus magnetic field at various temperatures. (e) The longitudinal resistivity versus magnetic field at various temperatures. Only the curve at 2 K is shifted downword vertically with step of 5$\mu\Omega\cdot cm$ for clarity. (f)(g) The first-order derivative of the curves in (d)(e), respectively. (h) The temperature-dependent anomalous Hall coefficient R$_{s}$. (i) Critical points at transitions picked from the first-order derivative of magnetization, Hall resistivity and longitudinal resistivity. The color of background only hints the trend of change without any other meaning.}
\end{figure*}

The high quality single crystals of Co$_{0.29}$TaS$_2$ were grown by chemical vapor transport method. The powder of Co, Ta and S with the ratio of 1:3:6 was sealed into a quartz tube before it was put into furnace and heated to 800$^\circ \mathrm{C}$ for 5 days to prepare polycrystalline precursor. The resulting powders (1g) and transport agent iodine (10 mg/cm$^3$) were then sealed in the quartz tube to grow single crystals with the temperature gradient set as 950$^\circ \mathrm{C}$ (source) and 850$^\circ \mathrm{C}$ (drain) for 10 days. Finally, single crystals with hexagonal plate shape were obtained, which are easily exfoliated. The atomic composition was confirmed to be Co:Ta:S = 0.29:1:2 by energy dispersive x-ray spectroscopy (EDS, Oxford X-Max 50). The single-crystal x-ray diffraction (XRD) and powder XRD patterns were obtained on a Brucker D8 Advance X-ray diffractometer using Cu K$_{\alpha}$ radiation. TOPAS-4.2 was employed for the refinement. The measurements of resistivity and magnetic properties were performed on the Quantum Design physical property measurement system (QD PPMS-14 T) and Quantum Design magnetic property measurement system (QD MPMS-3).

\section{Results and Discussion}

As shown in Fig. 1(a), the powder XRD patterns (powdered sample was obtained by crushing single crystals) are well refined with space group P6$_3$22, and the refined lattice parameters are a = b = 5.739 \AA \, and c = 11.912 \AA. The crystal structure is shown in the inset of Fig. 1(a), which consists of magnetic Co atoms intercalated between 2H-TaS$_2$ layers. The crystallographic c-axis is perpendicular to the plate, as shown in Fig. 1(b), and the side of crystal reveals the (11$\bar 2$0) crystalline surface. Inset of Fig. 1(b) displays a photo of Co$_{0.29}$TaS$_2$ single crystal with metallic luster. As displayed in Fig. 1(d), the temperature-dependent resistivity $\rho_{xx}$ demonstrates the metallic behavior with a sudden drop at around 38 K, which is due to the electron scattering as the onset of spin ordering.

The magnetic susceptibility of Co$_{0.29}$TaS$_2$ under the magnetic field 0.1 T applied along the c axis or within the ab plane is shown in the Fig. 2(a), respectively. Over the entire temperature range, the magnetic susceptibility along c axis $\chi_{c}$ exceeds that within ab plane $\chi_{ab}$, indicating that the easy axis is along the c direction. With decreasing temperature, $\chi_{c}$ exhibits a small peak centered at 38.3 K (labelled as T$_{t1}$), followed by a pronounced $\lambda$-shaped peak at 24.3 K (T$_{t2}$), both of which are consistent with the characteristics of AFM coupling. These two transitions are also discernible within ab plane. Above 50 K, the fit of average susceptibility $\chi_{ave}=(2/3)\chi_{ab}+(1/3)\chi_{c}$ to the modified Curie-Weiss law $\chi=\chi_{0}+C/(T-T_\theta)$ is presented in the inset of Fig. 2(a), where $\chi_{0}$ is the temperature-independent term resulting from the paramagnetism or diamagnetism, C stands for the Curie constant and T$_\theta$ represents the Weiss temperature. The negative value T$_{\theta}$ = -- 49 K indicates the predominant AFM interaction, and the obtained $\mu_{eff}=3.1 \mu_{B}$ per Co is smaller than the spin-only moment of 3.87$\mu_{B}$ for Co$^{2+}$ as reported in previous neutron scattering experiments\cite{Co0.33TaS}. The evolution of magnetic phase transition under different fields along c axis is observed from the ZFC susceptibility, as shown in Fig. 2(b). The corresponding first-order derivative is provided for clarity in the Fig. 2 (c). The transition temperature T$_{t2}$ decreases as the magnetic field increases, and T$_{t1}$ remains almost unchanged as the field changes. Meanwhile, another transition occurs at T$_{t3}\sim$ 34.3 K when the field $\sim$ 0.3 T is applied, which slightly decreases as the magnetic field increases, and disappears above 5 T.  

Figures 2(d) and (f) show the field-dependent magnetizations at various temperatures along c axis and within ab plane, respectively. As shown in the Fig. 2(d), except for the AFM background, a prominent hysteresis loop over the range of H$_{t1}\sim\pm6.9$ T at 2 K is observed, which is suppressed with increasing the temperature and disappears until T$_{t2}$. The hysteresis clearly reveals a tiny magnetic moment (0.0094 $\mu_{B}$ per Co atom) at zero field. Besides, there exists an obvious step around H$_{t2}\sim\pm$4 T, which becomes weaker as temperature increases and completely disappears near T$_{t1}$. These transitions are clearly displayed in the dM/dH versus $\mu_{0}$H plots as shown in the Fig. 2(e) in which the peaks evolve as the temperature changes. Meanwhile, when the magnetic field is applied within the ab plane, only linear magnetization curves are observed. These abundant temperature-dependent magnetic transitions indicate complex magnetic structures in Co$_{0.29}$TaS$_2$ which deserves detailed study, meanwhile, first-principles calculations and different kinds of spectroscopic methods are being employed by our collaborators to determine the possible magnetic structures.

Materials with abundant magnetic structures induce nontrivial transport responses usually demonstrated as anomalies in transverse Hall resistivity or longitudinal resistivity. The magnetization, Hall resistivity and longitudinal resistivity in Co$_{0.29}$TaS$_2$  are illustrated as a function of field H // c at 2 K in Figs. 3(a), (b) and (c). The pretty large Hall hysteresis and irreversible resistivity correspond to the hysteresis loop of magnetization over the range of H$_{t1}$. Simultaneously, the steps around H$_{t2}$ are consistent in magnetization, Hall and resistivity. Figure 3(d) shows the conspicuous AHE under zero field at various temperatures. Conventionally, in ferromagnet, the Hall resistivity is expressed as the combination of two terms,  $\rho_{yx}=\rho^{N}_{yx}+\rho^{A}_{yx}$, where $\rho^{N}_{yx}$ represents the ordinary Hall resistivity and $\rho^{A}_{yx}$ stands for the anomalous Hall resistivity. Further, the ordinary Hall resistivity $\rho^{N}_{yx}$ is linearly field-dependent while the anomalous term $\rho^{A}_{yx}$ is proportional to magnetization M, as follows,

\begin{equation}\label{equ1}
\centering
\rho_{yx}=R_{0}B+\mu_{0}R_{s}M
\end{equation}

\noindent where $R_{0}$ is the ordinary Hall coefficient and $\mu_{0}$ is the permeability, $R_{s}$ is the anomalous Hall coefficient and M is the magnetization. At 2 K, the ordinary Hall resistivity presents linear field dependence with positive slope between 5 T and 10 T, indicating that hole-type carriers dominate in Co$_{0.29}$TaS$_2$. With the fit of single band model, the ordinary Hall coefficient  R$_{0}\sim 2.3\times10^{-2}$ $\mu\Omega\cdot cm/T$, and thus the estimated carrier concentration n (n = $1/\arrowvert eR_{0}\arrowvert$) is 2.71$\times10^{22}$ $cm^{-3}$, where e is the charge of the electron. At 2 K, the large anomalous Hall resistivity $\rho^{A}_{yx}(H=0)$ obtained at zero field reaches 1.76 $\mu\Omega\cdot cm$ with the extremely small net magnetization $M(H=0) \sim$0.0094 $\mu_{B}$ per Co obtained at zero field. Accordingly, R$_{s}=\rho^{A}_{yx}(H=0)/\mu_{0}M(H=0)$ has significantly large value 3.40$\times10^{3}$ $\mu\Omega\cdot cm/T$, which is five orders of magnitude larger than R$_{0}$. As temperature increases, the value of R$_{s}$ increases slightly and reaches the maximum (4.07$\times10^{3}$ $\mu\Omega\cdot cm/T$) at 15 K as shown in Fig. 3(h), which diminishes rapidly when T$\geqslant$ T$_{t2}$ as ferromagnetic hysteresis loop disappears. Of particular interest is the extremely large ratio R$_{s}$/R$_{0}$ in Co$_{0.29}$TaS$_2$, particularly rare in antiferromagnetic materials except for Mn$_{3}$Sn\cite{Mn3Sn} and Co$_{x}$NbS$_2$\cite{CoNbS-1,CoNbS-2,CoNbS-3}, \textit{etc.} Moreover, the value is roughly two orders of magnitude larger than that of some ferromagnetic materials (about 100 $\sim$ 1000), such as Fe$_{0.25}$TaS$_{2}$\cite{Fe0.25TaS-2}, Co$_{3}$Sn$_{2}$S$_{2}$\cite{CoSnS-liu,CoSnS-wang}, Fe$_{3}$Sn$_{2}$\cite{Fe3Sn2} and Fe$_{3}$GeTe$_{2}$\cite{FeGeTe}. The exceedingly large AHE in 
Co$_{0.29}$TaS$_2$ is unlikely to be caused only by such tiny net magnetization. Therefore, there must exist other contributions to such large AHE, which are worthy to be investigated.

As a crucial measurement for inferring information about the interactions between carriers and the magnetic degrees of freedom in magnetic materials\cite{AMR}, magnetoresistivity results are presented below. The field-dependent resistivity measurements are performed on Co$_{0.29}$TaS$_2$ under various temperatures, as shown in the Fig. 3(e). The obvious irreversible behavior below T$_{t2}$ is consistent with the FM hysteresis loop with H applied along c axis. Noticed that the sharp change around H$_{t2}$ comes from the spin fluctuation around $\pm4$ T as displayed in the magnetization. The AFM coupling usually generates strong spin scattering, resulting in a high-resistivity state. While, the spin scattering is suppressed with FM coupling, corresponding to a low-resistivity state. Taking the magnetization and longitudinal resistivity curves at 2 K (Figs. 3(a) and 3(c)) as an example, the correlation between the magnetic transition and resistivity is analyzed. According to the history of applied field, the resistivity increases as the field decreases from +10 T to +5 T, then abruptly decreases until near the transition H$_{t2}$ (+4 T). The resistivity keeps increasing over the range of +4 T to -7 T, then begins to  increase sharply near the transition H$_{t1}$ (-7 T). Finally, the resistivity decreases further as the field decreases from -8 T to -10 T. Below -10 T (above +10 T), the longitudinal resistivity curves with increasing and decreasing fields coincide. In the boundary between the metamagnetic transitions at H$_{t1}$ and H$_{t2}$, sharp changes in resistivity are probably caused by spin reorientation. These behaviors reveal the complex magnetic structures, same as that indicated in Fig. 3(a). The transitions are suppressed by thermal fluctuation with the increase of the temperature. Based on the magnetization, Hall and longitudinal resistivity measurements, the phase diagram is plotted in Fig. 3(i). The same trend confirms the consistence of rich magnetic structures in the magnetization and magnetotransport.

\begin{figure}[htbp]
	\centering
	\includegraphics[width=0.48\textwidth]{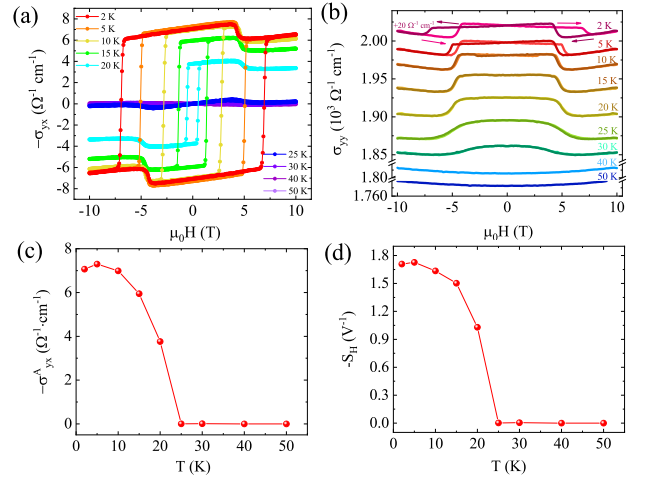}
	\caption{(a) The Hall conductivity as a function of magnetic field at various temperatures. (b) The longitudinal conductivity as a function of magnetic field at various temperatures. The curve at 2 K is shifted upward vertically with step of 20 $\Omega^{-1}\cdot cm^{-1}$ for clarity. (c) The anomalous Hall conductivity obtained in (a) at H=0. (d) The scaling coefficient -S$_{H}$ as a function of temperature.}
\end{figure}

The Hall conductivity $\sigma^{A}_{xy}$ and longitudinal conductivity $\sigma_{yy}$ are calculated by $-\sigma_{yx}=\rho_{yx}/(\rho^{2}_{yx}+\rho^{2}_{yy})$ and  $\sigma_{yy}=\rho_{yy}/(\rho^{2}_{yx}+\rho^{2}_{yy})$, respectively, which are plotted in the Figs. 4(a) and (b). The anomalous Hall conductivity $-\sigma^{A}_{yx} (H=0)$ versus T is plotted in the Fig. 4(c), which decreases as the temperature increases and vanishes above T$_{t1}$. The scaling coefficient S$_{H}=-\sigma^{A}_{yx} (H=0)/M(H=0)$ presents the same dependence of temperature, as shown in the Fig. 4(d). The derived value of S$_{H}$ $\sim$ 1.70 V$^{-1}$ at 2 K is exceedingly larger than that in traditional ferromagnets, such as Fe (0.06 V$^{-1}$)\cite{Fe} or Ni (-0.14 V$^{-1}$)\cite{Ni}.

\section{Summary}
In conclusion, we investigate the magnetic-atom-intercalated transition metal dichalcogenide Co$_{0.29}$TaS$_2$ and present its complex magnetic structure and novel magnetotransport. As the temperature decreases, Co$_{0.29}$TaS$_2$ successively experiences two magnetic transitions at T $_{t1}\sim $ 38.3 K and T$_{t2}\sim$ 24.3 K. With the magnetic field applied, over the range of 0.3 T - 5 T, another transition T$_{t3}\sim$ 34.3 K appears. A prominent hysteresis loop within H$_{t1}\sim\pm$ 6.9 T at 2 K is observed, which decreases with increasing temperature and disappears at T$_{t2}$. Besides, a step-like magnetic transition is detected in magnetization around H$_{t2}\sim\pm$4 T below T$_{t1}$. These magnetic transitions tuned by both temperature and magnetic field have a direct impact on the transport properties, which makes Co$_{0.29}$TaS$_2$ a fascinating 2D magnetic material. Strikingly, the AHE with extremely large value of R$_{s}$/R$_{0}\sim 10^{5}$ with tiny net magnetization (0.0094 $\mu_{B}$ per Co) is observed. The longitudinal resistivity shows a prominent irreversible behavior within H$_{t1}$. The abrupt change at H$_{t2}$ below T$_{t2}$, corresponding to the step-like magnetic transitions, is also observed in both Hall and longitudinal resistivity. The interplay between the magnetic structures and band topology in Co$_{0.29}$TaS$_2$ may be the key to explain the extremely large AHE in non-collinear antiferromagnet, which is in process with our collaborators. Co$_{0.29}$TaS$_2$ demonstrates complex but attractive magnetic structures, corresponding magnetotransport of which is worthy to explore, offering a new platform to study 2D magnetic materials. 

\emph{Note added.} When the paper is being prepared, we notice one related work posted on arXiv\cite{park2022field}, where similar transport results are reported and the AHE is attributed to hourglass Weyl fermion with the toroidal moment from the chiral lattice and geometric frustration.

\section{Acknowledgments}
We thank Yu Ye, Zhihai Cheng, Hongming Weng, Peng Cheng, Hang Li and Huancheng Yang for their fruitful discussions and collaborations. This work is supported by the National Key R $\&$ D Program of China (2019YFA0308602), the National Natural Science Foundation of China (Grants No. 12074425 and No. 11874422), and the Fundamental Research Funds for the Central Universities, and the Research Funds of Renmin University of China (No. 19XNLG18). Huan Wang is also supported by the Outstanding Innovative Talents Cultivation Funded Programs 2020 of Renmin University of China.

\normalem
\bibliography{Bibtex}

\begin{thebibliography}{39}
\expandafter\ifx\csname natexlab\endcsname\relax\def\natexlab#1{#1}\fi
\expandafter\ifx\csname bibnamefont\endcsname\relax
  \def\bibnamefont#1{#1}\fi
\expandafter\ifx\csname bibfnamefont\endcsname\relax
  \def\bibfnamefont#1{#1}\fi
\expandafter\ifx\csname citenamefont\endcsname\relax
  \def\citenamefont#1{#1}\fi
\expandafter\ifx\csname url\endcsname\relax
  \def\url#1{\texttt{#1}}\fi
\expandafter\ifx\csname urlprefix\endcsname\relax\def\urlprefix{URL }\fi
\providecommand{\bibinfo}[2]{#2}
\providecommand{\eprint}[2][]{\url{#2}}

\bibitem[{\citenamefont{Nagaosa et~al.}(2010)\citenamefont{Nagaosa, Sinova,
  Onoda, MacDonald, and Ong}}]{AHE}
\bibinfo{author}{\bibfnamefont{N.}~\bibnamefont{Nagaosa}},
  \bibinfo{author}{\bibfnamefont{J.}~\bibnamefont{Sinova}},
  \bibinfo{author}{\bibfnamefont{S.}~\bibnamefont{Onoda}},
  \bibinfo{author}{\bibfnamefont{A.~H.} \bibnamefont{MacDonald}},
  \bibnamefont{and} \bibinfo{author}{\bibfnamefont{N.~P.} \bibnamefont{Ong}},
  \bibinfo{journal}{Rev. Mod. Phys.} \textbf{\bibinfo{volume}{82}},
  \bibinfo{pages}{1539} (\bibinfo{year}{2010}).

\bibitem[{\citenamefont{Chen et~al.}(2014)\citenamefont{Chen, Niu, and
  MacDonald}}]{Mn3Ir-1}
\bibinfo{author}{\bibfnamefont{H.}~\bibnamefont{Chen}},
  \bibinfo{author}{\bibfnamefont{Q.}~\bibnamefont{Niu}}, \bibnamefont{and}
  \bibinfo{author}{\bibfnamefont{A.~H.} \bibnamefont{MacDonald}},
  \bibinfo{journal}{Phys. Rev. Lett.} \textbf{\bibinfo{volume}{112}},
  \bibinfo{pages}{017205} (\bibinfo{year}{2014}).

\bibitem[{\citenamefont{Suzuki et~al.}(2017)\citenamefont{Suzuki, Koretsune,
  Ochi, and Arita}}]{Mn3Ir-2}
\bibinfo{author}{\bibfnamefont{M.-T.} \bibnamefont{Suzuki}},
  \bibinfo{author}{\bibfnamefont{T.}~\bibnamefont{Koretsune}},
  \bibinfo{author}{\bibfnamefont{M.}~\bibnamefont{Ochi}}, \bibnamefont{and}
  \bibinfo{author}{\bibfnamefont{R.}~\bibnamefont{Arita}},
  \bibinfo{journal}{Phys. Rev. B} \textbf{\bibinfo{volume}{95}},
  \bibinfo{pages}{094406} (\bibinfo{year}{2017}).

\bibitem[{\citenamefont{Shindou and Nagaosa}(2001)}]{AFM-1}
\bibinfo{author}{\bibfnamefont{R.}~\bibnamefont{Shindou}} \bibnamefont{and}
  \bibinfo{author}{\bibfnamefont{N.}~\bibnamefont{Nagaosa}},
  \bibinfo{journal}{Phys. Rev. Lett.} \textbf{\bibinfo{volume}{87}},
  \bibinfo{pages}{116801} (\bibinfo{year}{2001}).

\bibitem[{\citenamefont{K{\"u}bler and Felser}(2014)}]{AFM-2}
\bibinfo{author}{\bibfnamefont{J.}~\bibnamefont{K{\"u}bler}} \bibnamefont{and}
  \bibinfo{author}{\bibfnamefont{C.}~\bibnamefont{Felser}},
  \bibinfo{journal}{EPL} \textbf{\bibinfo{volume}{108}}, \bibinfo{pages}{67001}
  (\bibinfo{year}{2014}).

\bibitem[{\citenamefont{Smejkal et~al.}(2021)\citenamefont{Smejkal, MacDonald,
  Sinova, Nakatsuji, and Jungwirth}}]{AFM-3}
\bibinfo{author}{\bibfnamefont{L.}~\bibnamefont{Smejkal}},
  \bibinfo{author}{\bibfnamefont{A.~H.} \bibnamefont{MacDonald}},
  \bibinfo{author}{\bibfnamefont{J.}~\bibnamefont{Sinova}},
  \bibinfo{author}{\bibfnamefont{S.}~\bibnamefont{Nakatsuji}},
  \bibnamefont{and}
  \bibinfo{author}{\bibfnamefont{T.}~\bibnamefont{Jungwirth}},
  \bibinfo{journal}{arXiv:2107.03321}  (\bibinfo{year}{2021}).

\bibitem[{\citenamefont{Bonbien et~al.}(2021)\citenamefont{Bonbien, Zhuo,
  Salimath, Ly, Abbout, and Manchon}}]{AFM-4}
\bibinfo{author}{\bibfnamefont{V.}~\bibnamefont{Bonbien}},
  \bibinfo{author}{\bibfnamefont{F.}~\bibnamefont{Zhuo}},
  \bibinfo{author}{\bibfnamefont{A.}~\bibnamefont{Salimath}},
  \bibinfo{author}{\bibfnamefont{O.}~\bibnamefont{Ly}},
  \bibinfo{author}{\bibfnamefont{A.}~\bibnamefont{Abbout}}, \bibnamefont{and}
  \bibinfo{author}{\bibfnamefont{A.}~\bibnamefont{Manchon}},
  \bibinfo{journal}{J. Phys. D: Appl. Phy.} \textbf{\bibinfo{volume}{55}},
  \bibinfo{pages}{103002} (\bibinfo{year}{2021}).

\bibitem[{\citenamefont{Nayak et~al.}(2016)\citenamefont{Nayak, Fischer, Sun,
  Yan, Karel, Komarek, Shekhar, Kumar, Schnelle, K{\"u}bler et~al.}}]{Mn3Ge}
\bibinfo{author}{\bibfnamefont{A.~K.} \bibnamefont{Nayak}},
  \bibinfo{author}{\bibfnamefont{J.~E.} \bibnamefont{Fischer}},
  \bibinfo{author}{\bibfnamefont{Y.}~\bibnamefont{Sun}},
  \bibinfo{author}{\bibfnamefont{B.}~\bibnamefont{Yan}},
  \bibinfo{author}{\bibfnamefont{J.}~\bibnamefont{Karel}},
  \bibinfo{author}{\bibfnamefont{A.~C.} \bibnamefont{Komarek}},
  \bibinfo{author}{\bibfnamefont{C.}~\bibnamefont{Shekhar}},
  \bibinfo{author}{\bibfnamefont{N.}~\bibnamefont{Kumar}},
  \bibinfo{author}{\bibfnamefont{W.}~\bibnamefont{Schnelle}},
  \bibinfo{author}{\bibfnamefont{J.}~\bibnamefont{K{\"u}bler}},
  \bibnamefont{et~al.}, \bibinfo{journal}{Sci. Adv}
  \textbf{\bibinfo{volume}{2}}, \bibinfo{pages}{e1501870}
  (\bibinfo{year}{2016}).

\bibitem[{\citenamefont{Nakatsuji et~al.}(2015)\citenamefont{Nakatsuji,
  Kiyohara, and Higo}}]{Mn3Sn}
\bibinfo{author}{\bibfnamefont{S.}~\bibnamefont{Nakatsuji}},
  \bibinfo{author}{\bibfnamefont{N.}~\bibnamefont{Kiyohara}}, \bibnamefont{and}
  \bibinfo{author}{\bibfnamefont{T.}~\bibnamefont{Higo}},
  \bibinfo{journal}{Nature} \textbf{\bibinfo{volume}{527}},
  \bibinfo{pages}{212} (\bibinfo{year}{2015}).

\bibitem[{\citenamefont{Chen et~al.}(2021)\citenamefont{Chen, Tomita, Minami,
  Fu, Koretsune, Kitatani, Muhammad, Nishio-Hamane, Ishii, Ishii
  et~al.}}]{Mn3X}
\bibinfo{author}{\bibfnamefont{T.}~\bibnamefont{Chen}},
  \bibinfo{author}{\bibfnamefont{T.}~\bibnamefont{Tomita}},
  \bibinfo{author}{\bibfnamefont{S.}~\bibnamefont{Minami}},
  \bibinfo{author}{\bibfnamefont{M.}~\bibnamefont{Fu}},
  \bibinfo{author}{\bibfnamefont{T.}~\bibnamefont{Koretsune}},
  \bibinfo{author}{\bibfnamefont{M.}~\bibnamefont{Kitatani}},
  \bibinfo{author}{\bibfnamefont{I.}~\bibnamefont{Muhammad}},
  \bibinfo{author}{\bibfnamefont{D.}~\bibnamefont{Nishio-Hamane}},
  \bibinfo{author}{\bibfnamefont{R.}~\bibnamefont{Ishii}},
  \bibinfo{author}{\bibfnamefont{F.}~\bibnamefont{Ishii}},
  \bibnamefont{et~al.}, \bibinfo{journal}{Nat.Commun.}
  \textbf{\bibinfo{volume}{12}}, \bibinfo{pages}{572} (\bibinfo{year}{2021}).

\bibitem[{\citenamefont{S{\"u}rgers et~al.}(2017)\citenamefont{S{\"u}rgers,
  Wolf, Adelmann, Kittler, Fischer, and L{\"o}hneysen}}]{Mn5Si3}
\bibinfo{author}{\bibfnamefont{C.}~\bibnamefont{S{\"u}rgers}},
  \bibinfo{author}{\bibfnamefont{T.}~\bibnamefont{Wolf}},
  \bibinfo{author}{\bibfnamefont{P.}~\bibnamefont{Adelmann}},
  \bibinfo{author}{\bibfnamefont{W.}~\bibnamefont{Kittler}},
  \bibinfo{author}{\bibfnamefont{G.}~\bibnamefont{Fischer}}, \bibnamefont{and}
  \bibinfo{author}{\bibfnamefont{H.~v.} \bibnamefont{L{\"o}hneysen}},
  \bibinfo{journal}{Sci. Rep.} \textbf{\bibinfo{volume}{7}},
  \bibinfo{pages}{42982} (\bibinfo{year}{2017}).

\bibitem[{\citenamefont{Suzuki et~al.}(2016)\citenamefont{Suzuki, Chisnell,
  Devarakonda, Liu, Feng, Xiao, Lynn, and Checkelsky}}]{GdPtBi}
\bibinfo{author}{\bibfnamefont{T.}~\bibnamefont{Suzuki}},
  \bibinfo{author}{\bibfnamefont{R.}~\bibnamefont{Chisnell}},
  \bibinfo{author}{\bibfnamefont{A.}~\bibnamefont{Devarakonda}},
  \bibinfo{author}{\bibfnamefont{Y.~T.} \bibnamefont{Liu}},
  \bibinfo{author}{\bibfnamefont{W.}~\bibnamefont{Feng}},
  \bibinfo{author}{\bibfnamefont{D.}~\bibnamefont{Xiao}},
  \bibinfo{author}{\bibfnamefont{J.~W.} \bibnamefont{Lynn}}, \bibnamefont{and}
  \bibinfo{author}{\bibfnamefont{J.~G.} \bibnamefont{Checkelsky}},
  \bibinfo{journal}{Nature Phys.} \textbf{\bibinfo{volume}{12}},
  \bibinfo{pages}{1119} (\bibinfo{year}{2016}).

\bibitem[{\citenamefont{Taguchi et~al.}(2001)\citenamefont{Taguchi, Oohara,
  Yoshizawa, Nagaosa, and Tokura}}]{NdMoO}
\bibinfo{author}{\bibfnamefont{Y.}~\bibnamefont{Taguchi}},
  \bibinfo{author}{\bibfnamefont{Y.}~\bibnamefont{Oohara}},
  \bibinfo{author}{\bibfnamefont{H.}~\bibnamefont{Yoshizawa}},
  \bibinfo{author}{\bibfnamefont{N.}~\bibnamefont{Nagaosa}}, \bibnamefont{and}
  \bibinfo{author}{\bibfnamefont{Y.}~\bibnamefont{Tokura}},
  \bibinfo{journal}{Science} \textbf{\bibinfo{volume}{291}},
  \bibinfo{pages}{2573} (\bibinfo{year}{2001}).

\bibitem[{\citenamefont{Machida et~al.}(2010)\citenamefont{Machida, Nakatsuji,
  Onoda, Tayama, and Sakakibara}}]{PrIrO}
\bibinfo{author}{\bibfnamefont{Y.}~\bibnamefont{Machida}},
  \bibinfo{author}{\bibfnamefont{S.}~\bibnamefont{Nakatsuji}},
  \bibinfo{author}{\bibfnamefont{S.}~\bibnamefont{Onoda}},
  \bibinfo{author}{\bibfnamefont{T.}~\bibnamefont{Tayama}}, \bibnamefont{and}
  \bibinfo{author}{\bibfnamefont{T.}~\bibnamefont{Sakakibara}},
  \bibinfo{journal}{Nature} \textbf{\bibinfo{volume}{463}},
  \bibinfo{pages}{210} (\bibinfo{year}{2010}).

\bibitem[{\citenamefont{Parkin and Friend}(1980{\natexlab{a}})}]{series-1}
\bibinfo{author}{\bibfnamefont{S.~S.~P.} \bibnamefont{Parkin}}
  \bibnamefont{and} \bibinfo{author}{\bibfnamefont{R.~H.}
  \bibnamefont{Friend}}, \bibinfo{journal}{Philos. Mag.}
  \textbf{\bibinfo{volume}{41}}, \bibinfo{pages}{65}
  (\bibinfo{year}{1980}{\natexlab{a}}).

\bibitem[{\citenamefont{Parkin and Friend}(1980{\natexlab{b}})}]{series-2}
\bibinfo{author}{\bibfnamefont{S.~S.~P.} \bibnamefont{Parkin}}
  \bibnamefont{and} \bibinfo{author}{\bibfnamefont{R.~H.}
  \bibnamefont{Friend}}, \bibinfo{journal}{Philos. Mag.}
  \textbf{\bibinfo{volume}{41}}, \bibinfo{pages}{95}
  (\bibinfo{year}{1980}{\natexlab{b}}).

\bibitem[{\citenamefont{Lu et~al.}(2020)\citenamefont{Lu, Sapkota,
  DeBeer-Schmitt, Wu, Cao, Mannella, Mandrus, Aczel, and MacDougall}}]{VTa3S6}
\bibinfo{author}{\bibfnamefont{K.}~\bibnamefont{Lu}},
  \bibinfo{author}{\bibfnamefont{D.}~\bibnamefont{Sapkota}},
  \bibinfo{author}{\bibfnamefont{L.}~\bibnamefont{DeBeer-Schmitt}},
  \bibinfo{author}{\bibfnamefont{Y.}~\bibnamefont{Wu}},
  \bibinfo{author}{\bibfnamefont{H.}~\bibnamefont{Cao}},
  \bibinfo{author}{\bibfnamefont{N.}~\bibnamefont{Mannella}},
  \bibinfo{author}{\bibfnamefont{D.}~\bibnamefont{Mandrus}},
  \bibinfo{author}{\bibfnamefont{A.~A.} \bibnamefont{Aczel}}, \bibnamefont{and}
  \bibinfo{author}{\bibfnamefont{G.~J.} \bibnamefont{MacDougall}},
  \bibinfo{journal}{Phys. Rev. Mater.} \textbf{\bibinfo{volume}{4}},
  \bibinfo{pages}{054416} (\bibinfo{year}{2020}).

\bibitem[{\citenamefont{Yamasaki et~al.}(2017)\citenamefont{Yamasaki, Moriya,
  Arai, Masubuchi, Pyon, Tamegai, Ueno, and Machida}}]{CrTa3S6-1}
\bibinfo{author}{\bibfnamefont{Y.}~\bibnamefont{Yamasaki}},
  \bibinfo{author}{\bibfnamefont{R.}~\bibnamefont{Moriya}},
  \bibinfo{author}{\bibfnamefont{M.}~\bibnamefont{Arai}},
  \bibinfo{author}{\bibfnamefont{S.}~\bibnamefont{Masubuchi}},
  \bibinfo{author}{\bibfnamefont{S.}~\bibnamefont{Pyon}},
  \bibinfo{author}{\bibfnamefont{T.}~\bibnamefont{Tamegai}},
  \bibinfo{author}{\bibfnamefont{K.}~\bibnamefont{Ueno}}, \bibnamefont{and}
  \bibinfo{author}{\bibfnamefont{T.}~\bibnamefont{Machida}},
  \bibinfo{journal}{2D Mater.} \textbf{\bibinfo{volume}{4}},
  \bibinfo{pages}{041007} (\bibinfo{year}{2017}).

\bibitem[{\citenamefont{Kousaka et~al.}(2016)\citenamefont{Kousaka, Ogura,
  Zhang, Miao, Lee, Torii, Kamiyama, Campo, Inoue, and Akimitsu}}]{CrTa3S6-2}
\bibinfo{author}{\bibfnamefont{Y.}~\bibnamefont{Kousaka}},
  \bibinfo{author}{\bibfnamefont{T.}~\bibnamefont{Ogura}},
  \bibinfo{author}{\bibfnamefont{J.}~\bibnamefont{Zhang}},
  \bibinfo{author}{\bibfnamefont{P.}~\bibnamefont{Miao}},
  \bibinfo{author}{\bibfnamefont{S.}~\bibnamefont{Lee}},
  \bibinfo{author}{\bibfnamefont{S.}~\bibnamefont{Torii}},
  \bibinfo{author}{\bibfnamefont{T.}~\bibnamefont{Kamiyama}},
  \bibinfo{author}{\bibfnamefont{J.}~\bibnamefont{Campo}},
  \bibinfo{author}{\bibfnamefont{K.}~\bibnamefont{Inoue}}, \bibnamefont{and}
  \bibinfo{author}{\bibfnamefont{J.}~\bibnamefont{Akimitsu}}, in
  \emph{\bibinfo{booktitle}{J.Phys.Conf.Ser.}} (\bibinfo{organization}{IOP
  Publishing}, \bibinfo{year}{2016}), vol. \bibinfo{volume}{746}, p.
  \bibinfo{pages}{012061}.

\bibitem[{\citenamefont{Zhang et~al.}(2018)\citenamefont{Zhang, Wei, Zheng, Lu,
  Wu, Zhu, Tang, Ning, Han, Ling et~al.}}]{MnTa3S6}
\bibinfo{author}{\bibfnamefont{H.}~\bibnamefont{Zhang}},
  \bibinfo{author}{\bibfnamefont{W.}~\bibnamefont{Wei}},
  \bibinfo{author}{\bibfnamefont{G.}~\bibnamefont{Zheng}},
  \bibinfo{author}{\bibfnamefont{J.}~\bibnamefont{Lu}},
  \bibinfo{author}{\bibfnamefont{M.}~\bibnamefont{Wu}},
  \bibinfo{author}{\bibfnamefont{X.}~\bibnamefont{Zhu}},
  \bibinfo{author}{\bibfnamefont{J.}~\bibnamefont{Tang}},
  \bibinfo{author}{\bibfnamefont{W.}~\bibnamefont{Ning}},
  \bibinfo{author}{\bibfnamefont{Y.}~\bibnamefont{Han}},
  \bibinfo{author}{\bibfnamefont{L.}~\bibnamefont{Ling}}, \bibnamefont{et~al.},
  \bibinfo{journal}{Appl.Phys.Lett.} \textbf{\bibinfo{volume}{113}},
  \bibinfo{pages}{072402} (\bibinfo{year}{2018}).

\bibitem[{\citenamefont{Liang}(2017)}]{MnTa4S8}
\bibinfo{author}{\bibfnamefont{L.}~\bibnamefont{Liang}}, Ph.D. thesis,
  \bibinfo{school}{University of Groningen} (\bibinfo{year}{2017}).

\bibitem[{\citenamefont{Liu et~al.}(2021)\citenamefont{Liu, Hu, Stavitski,
  Attenkofer, Petrovic et~al.}}]{Co0.22TaS}
\bibinfo{author}{\bibfnamefont{Y.}~\bibnamefont{Liu}},
  \bibinfo{author}{\bibfnamefont{Z.}~\bibnamefont{Hu}},
  \bibinfo{author}{\bibfnamefont{E.}~\bibnamefont{Stavitski}},
  \bibinfo{author}{\bibfnamefont{K.}~\bibnamefont{Attenkofer}},
  \bibinfo{author}{\bibfnamefont{C.}~\bibnamefont{Petrovic}},
  \bibnamefont{et~al.}, \bibinfo{journal}{Phys.Rev.Res.}
  \textbf{\bibinfo{volume}{3}}, \bibinfo{pages}{023181} (\bibinfo{year}{2021}).

\bibitem[{\citenamefont{Parkin et~al.}(1983)\citenamefont{Parkin, Marseglia,
  and Brown}}]{Co0.33TaS}
\bibinfo{author}{\bibfnamefont{S.~S.~P.} \bibnamefont{Parkin}},
  \bibinfo{author}{\bibfnamefont{E.~A.} \bibnamefont{Marseglia}},
  \bibnamefont{and} \bibinfo{author}{\bibfnamefont{P.~J.} \bibnamefont{Brown}},
  \bibinfo{journal}{J. Phys. C: Solid State Phys.}
  \textbf{\bibinfo{volume}{16}}, \bibinfo{pages}{2765} (\bibinfo{year}{1983}).

\bibitem[{\citenamefont{Morosan et~al.}(2007)\citenamefont{Morosan, Zandbergen,
  Li, Lee, Checkelsky, Heinrich, Siegrist, Ong, and Cava}}]{Fe0.25TaS-1}
\bibinfo{author}{\bibfnamefont{E.}~\bibnamefont{Morosan}},
  \bibinfo{author}{\bibfnamefont{H.~W.} \bibnamefont{Zandbergen}},
  \bibinfo{author}{\bibfnamefont{L.}~\bibnamefont{Li}},
  \bibinfo{author}{\bibfnamefont{M.}~\bibnamefont{Lee}},
  \bibinfo{author}{\bibfnamefont{J.~G.} \bibnamefont{Checkelsky}},
  \bibinfo{author}{\bibfnamefont{M.}~\bibnamefont{Heinrich}},
  \bibinfo{author}{\bibfnamefont{T.}~\bibnamefont{Siegrist}},
  \bibinfo{author}{\bibfnamefont{N.~P.} \bibnamefont{Ong}}, \bibnamefont{and}
  \bibinfo{author}{\bibfnamefont{R.~J.} \bibnamefont{Cava}},
  \bibinfo{journal}{Phys. Rev. B} \textbf{\bibinfo{volume}{75}},
  \bibinfo{pages}{104401} (\bibinfo{year}{2007}).

\bibitem[{\citenamefont{Checkelsky et~al.}(2008)\citenamefont{Checkelsky, Lee,
  Morosan, Cava, and Ong}}]{Fe0.25TaS-2}
\bibinfo{author}{\bibfnamefont{J.~G.} \bibnamefont{Checkelsky}},
  \bibinfo{author}{\bibfnamefont{M.}~\bibnamefont{Lee}},
  \bibinfo{author}{\bibfnamefont{E.}~\bibnamefont{Morosan}},
  \bibinfo{author}{\bibfnamefont{R.~J.} \bibnamefont{Cava}}, \bibnamefont{and}
  \bibinfo{author}{\bibfnamefont{N.~P.} \bibnamefont{Ong}},
  \bibinfo{journal}{Phys. Rev. B} \textbf{\bibinfo{volume}{77}},
  \bibinfo{pages}{014433} (\bibinfo{year}{2008}).

\bibitem[{\citenamefont{Hardy et~al.}(2015)\citenamefont{Hardy, Chen,
  Marcinkova, Ji, Sinova, Natelson, and Morosan}}]{Fe0.28TaS-1}
\bibinfo{author}{\bibfnamefont{W.~J.} \bibnamefont{Hardy}},
  \bibinfo{author}{\bibfnamefont{C.-W.} \bibnamefont{Chen}},
  \bibinfo{author}{\bibfnamefont{A.}~\bibnamefont{Marcinkova}},
  \bibinfo{author}{\bibfnamefont{H.}~\bibnamefont{Ji}},
  \bibinfo{author}{\bibfnamefont{J.}~\bibnamefont{Sinova}},
  \bibinfo{author}{\bibfnamefont{D.}~\bibnamefont{Natelson}}, \bibnamefont{and}
  \bibinfo{author}{\bibfnamefont{E.}~\bibnamefont{Morosan}},
  \bibinfo{journal}{Phys.Rev.B} \textbf{\bibinfo{volume}{91}},
  \bibinfo{pages}{054426} (\bibinfo{year}{2015}).

\bibitem[{\citenamefont{Dijkstra et~al.}(1989)\citenamefont{Dijkstra, Zijlema,
  Van~Bruggen, Haas, and de~Groot}}]{Fe0.28TaS-2}
\bibinfo{author}{\bibfnamefont{J.}~\bibnamefont{Dijkstra}},
  \bibinfo{author}{\bibfnamefont{P.~J.} \bibnamefont{Zijlema}},
  \bibinfo{author}{\bibfnamefont{C.~F.} \bibnamefont{Van~Bruggen}},
  \bibinfo{author}{\bibfnamefont{C.}~\bibnamefont{Haas}}, \bibnamefont{and}
  \bibinfo{author}{\bibfnamefont{R.~A.} \bibnamefont{de~Groot}},
  \bibinfo{journal}{J.Phys.Condens.Matter} \textbf{\bibinfo{volume}{1}},
  \bibinfo{pages}{6363} (\bibinfo{year}{1989}).

\bibitem[{\citenamefont{Narita et~al.}(1994)\citenamefont{Narita, Ikuta,
  Hinode, Uchida, Ohtani, and Wakihara}}]{FexTaS-AFM}
\bibinfo{author}{\bibfnamefont{H.}~\bibnamefont{Narita}},
  \bibinfo{author}{\bibfnamefont{H.}~\bibnamefont{Ikuta}},
  \bibinfo{author}{\bibfnamefont{H.}~\bibnamefont{Hinode}},
  \bibinfo{author}{\bibfnamefont{T.}~\bibnamefont{Uchida}},
  \bibinfo{author}{\bibfnamefont{T.}~\bibnamefont{Ohtani}}, \bibnamefont{and}
  \bibinfo{author}{\bibfnamefont{M.}~\bibnamefont{Wakihara}},
  \bibinfo{journal}{J. Solid State Chem.} \textbf{\bibinfo{volume}{108}},
  \bibinfo{pages}{148} (\bibinfo{year}{1994}).

\bibitem[{\citenamefont{Ghimire et~al.}(2018)\citenamefont{Ghimire, Botana,
  Jiang, Zhang, Chen, and Mitchell}}]{CoNbS-1}
\bibinfo{author}{\bibfnamefont{N.~J.} \bibnamefont{Ghimire}},
  \bibinfo{author}{\bibfnamefont{A.}~\bibnamefont{Botana}},
  \bibinfo{author}{\bibfnamefont{J.}~\bibnamefont{Jiang}},
  \bibinfo{author}{\bibfnamefont{J.}~\bibnamefont{Zhang}},
  \bibinfo{author}{\bibfnamefont{Y.-S.} \bibnamefont{Chen}}, \bibnamefont{and}
  \bibinfo{author}{\bibfnamefont{J.}~\bibnamefont{Mitchell}},
  \bibinfo{journal}{Nat.Commun.} \textbf{\bibinfo{volume}{9}},
  \bibinfo{pages}{1} (\bibinfo{year}{2018}).

\bibitem[{\citenamefont{Tenasini et~al.}(2020)\citenamefont{Tenasini, Martino,
  Ubrig, Ghimire, Berger, Zaharko, Wu, Mitchell, Martin, Forr{\'o}
  et~al.}}]{CoNbS-2}
\bibinfo{author}{\bibfnamefont{G.}~\bibnamefont{Tenasini}},
  \bibinfo{author}{\bibfnamefont{E.}~\bibnamefont{Martino}},
  \bibinfo{author}{\bibfnamefont{N.}~\bibnamefont{Ubrig}},
  \bibinfo{author}{\bibfnamefont{N.~J.} \bibnamefont{Ghimire}},
  \bibinfo{author}{\bibfnamefont{H.}~\bibnamefont{Berger}},
  \bibinfo{author}{\bibfnamefont{O.}~\bibnamefont{Zaharko}},
  \bibinfo{author}{\bibfnamefont{F.}~\bibnamefont{Wu}},
  \bibinfo{author}{\bibfnamefont{J.}~\bibnamefont{Mitchell}},
  \bibinfo{author}{\bibfnamefont{I.}~\bibnamefont{Martin}},
  \bibinfo{author}{\bibfnamefont{L.}~\bibnamefont{Forr{\'o}}},
  \bibnamefont{et~al.}, \bibinfo{journal}{Phys.Rev.Res.}
  \textbf{\bibinfo{volume}{2}}, \bibinfo{pages}{023051} (\bibinfo{year}{2020}).

\bibitem[{\citenamefont{Mangelsen et~al.}(2021)\citenamefont{Mangelsen, Zimmer,
  N{\"a}ther, Mankovsky, Polesya, Ebert, and Bensch}}]{CoNbS-3}
\bibinfo{author}{\bibfnamefont{S.}~\bibnamefont{Mangelsen}},
  \bibinfo{author}{\bibfnamefont{P.}~\bibnamefont{Zimmer}},
  \bibinfo{author}{\bibfnamefont{C.}~\bibnamefont{N{\"a}ther}},
  \bibinfo{author}{\bibfnamefont{S.}~\bibnamefont{Mankovsky}},
  \bibinfo{author}{\bibfnamefont{S.}~\bibnamefont{Polesya}},
  \bibinfo{author}{\bibfnamefont{H.}~\bibnamefont{Ebert}}, \bibnamefont{and}
  \bibinfo{author}{\bibfnamefont{W.}~\bibnamefont{Bensch}},
  \bibinfo{journal}{Phys. Rev. B} \textbf{\bibinfo{volume}{103}},
  \bibinfo{pages}{184408} (\bibinfo{year}{2021}).

\bibitem[{\citenamefont{Liu et~al.}(2018)\citenamefont{Liu, Sun, Kumar,
  Muechler, Sun, Jiao, Yang, Liu, Liang, Xu et~al.}}]{CoSnS-liu}
\bibinfo{author}{\bibfnamefont{E.}~\bibnamefont{Liu}},
  \bibinfo{author}{\bibfnamefont{Y.}~\bibnamefont{Sun}},
  \bibinfo{author}{\bibfnamefont{N.}~\bibnamefont{Kumar}},
  \bibinfo{author}{\bibfnamefont{L.}~\bibnamefont{Muechler}},
  \bibinfo{author}{\bibfnamefont{A.}~\bibnamefont{Sun}},
  \bibinfo{author}{\bibfnamefont{L.}~\bibnamefont{Jiao}},
  \bibinfo{author}{\bibfnamefont{S.-Y.} \bibnamefont{Yang}},
  \bibinfo{author}{\bibfnamefont{D.}~\bibnamefont{Liu}},
  \bibinfo{author}{\bibfnamefont{A.}~\bibnamefont{Liang}},
  \bibinfo{author}{\bibfnamefont{Q.}~\bibnamefont{Xu}}, \bibnamefont{et~al.},
  \bibinfo{journal}{Nat. Phys.} \textbf{\bibinfo{volume}{14}},
  \bibinfo{pages}{1125} (\bibinfo{year}{2018}).

\bibitem[{\citenamefont{Wang et~al.}(2018)\citenamefont{Wang, Xu, Lou, Liu, Li,
  Huang, Shen, Weng, Wang, and Lei}}]{CoSnS-wang}
\bibinfo{author}{\bibfnamefont{Q.}~\bibnamefont{Wang}},
  \bibinfo{author}{\bibfnamefont{Y.}~\bibnamefont{Xu}},
  \bibinfo{author}{\bibfnamefont{R.}~\bibnamefont{Lou}},
  \bibinfo{author}{\bibfnamefont{Z.}~\bibnamefont{Liu}},
  \bibinfo{author}{\bibfnamefont{M.}~\bibnamefont{Li}},
  \bibinfo{author}{\bibfnamefont{Y.}~\bibnamefont{Huang}},
  \bibinfo{author}{\bibfnamefont{D.}~\bibnamefont{Shen}},
  \bibinfo{author}{\bibfnamefont{H.}~\bibnamefont{Weng}},
  \bibinfo{author}{\bibfnamefont{S.}~\bibnamefont{Wang}}, \bibnamefont{and}
  \bibinfo{author}{\bibfnamefont{H.}~\bibnamefont{Lei}}, \bibinfo{journal}{Nat.
  Commun.} \textbf{\bibinfo{volume}{9}}, \bibinfo{pages}{3681}
  (\bibinfo{year}{2018}).

\bibitem[{\citenamefont{Qi et~al.}(2016)\citenamefont{Qi, Sun, Xiao, Fei, and
  Lei}}]{Fe3Sn2}
\bibinfo{author}{\bibfnamefont{W.}~\bibnamefont{Qi}},
  \bibinfo{author}{\bibfnamefont{S.}~\bibnamefont{Sun}},
  \bibinfo{author}{\bibfnamefont{Z.}~\bibnamefont{Xiao}},
  \bibinfo{author}{\bibfnamefont{P.}~\bibnamefont{Fei}}, \bibnamefont{and}
  \bibinfo{author}{\bibfnamefont{H.}~\bibnamefont{Lei}},
  \bibinfo{journal}{Phys. Rev. B} \textbf{\bibinfo{volume}{94}},
  \bibinfo{pages}{075135} (\bibinfo{year}{2016}).

\bibitem[{\citenamefont{Wang et~al.}(2017)\citenamefont{Wang, Cong, Jian, Liu,
  and Xiong}}]{FeGeTe}
\bibinfo{author}{\bibfnamefont{Y.}~\bibnamefont{Wang}},
  \bibinfo{author}{\bibfnamefont{X.}~\bibnamefont{Cong}},
  \bibinfo{author}{\bibfnamefont{W.}~\bibnamefont{Jian}},
  \bibinfo{author}{\bibfnamefont{B.}~\bibnamefont{Liu}}, \bibnamefont{and}
  \bibinfo{author}{\bibfnamefont{Y.}~\bibnamefont{Xiong}},
  \bibinfo{journal}{Phys. Rev. B} \textbf{\bibinfo{volume}{96}},
  \bibinfo{pages}{134428} (\bibinfo{year}{2017}).

\bibitem[{\citenamefont{McGuire and Potter}(1975)}]{AMR}
\bibinfo{author}{\bibfnamefont{T.}~\bibnamefont{McGuire}} \bibnamefont{and}
  \bibinfo{author}{\bibfnamefont{R.}~\bibnamefont{Potter}},
  \bibinfo{journal}{IEEE Trans. Magn.} \textbf{\bibinfo{volume}{11}},
  \bibinfo{pages}{1018} (\bibinfo{year}{1975}).

\bibitem[{\citenamefont{Dheer and P.}(1967)}]{Fe}
\bibinfo{author}{\bibnamefont{Dheer}} \bibnamefont{and}
  \bibinfo{author}{\bibfnamefont{N.}~\bibnamefont{P.}}, \bibinfo{journal}{Phys.
  Rev.} \textbf{\bibinfo{volume}{156}}, \bibinfo{pages}{637}
  (\bibinfo{year}{1967}).

\bibitem[{\citenamefont{Jan and Gijsman}(1952)}]{Ni}
\bibinfo{author}{\bibfnamefont{J.-P.} \bibnamefont{Jan}} \bibnamefont{and}
  \bibinfo{author}{\bibfnamefont{H.}~\bibnamefont{Gijsman}},
  \bibinfo{journal}{Physica} \textbf{\bibinfo{volume}{18}},
  \bibinfo{pages}{339} (\bibinfo{year}{1952}).

\bibitem[{\citenamefont{Park et~al.}(2022)\citenamefont{Park, Kang, Kim, Lee,
  Noh, Han, and Park}}]{park2022field}
\bibinfo{author}{\bibfnamefont{P.}~\bibnamefont{Park}},
  \bibinfo{author}{\bibfnamefont{Y.-G.} \bibnamefont{Kang}},
  \bibinfo{author}{\bibfnamefont{J.}~\bibnamefont{Kim}},
  \bibinfo{author}{\bibfnamefont{K.~H.} \bibnamefont{Lee}},
  \bibinfo{author}{\bibfnamefont{H.-J.} \bibnamefont{Noh}},
  \bibinfo{author}{\bibfnamefont{M.~J.} \bibnamefont{Han}}, \bibnamefont{and}
  \bibinfo{author}{\bibfnamefont{J.-G.} \bibnamefont{Park}},
  \bibinfo{journal}{arXiv:2203.03826}  (\bibinfo{year}{2022}).

\end{thebibliography}
\end{document}